# The Influence of Social Movements on Space Astronomy Policy

**The cases of "Hubble Huggers", JWST's "Science Warriors" and the ISEE-3 "Reboot Team"**

Hannah E. Harris[a,b] & Pedro Russo[a]

[a]Leiden University, the Netherlands, E-mail: russo@strw.leidenuniv.nl
[b]Wellesley College, the United States

## Abstract

Public engagement (PE) initiatives can lead to a long term public support of science. However most of the real impact of PE initiatives within the context of long-term science policy is not completely understood. An examination of the National Aeronautics and Space Administration's (NASA) Hubble Space Telescope, James Webb Space Telescope, and International Sun/Earth Explorer 3 reveal how large grassroots movements led by citizen scientists and space aficionados can have profound effects on public policy. We explore the role and relevance of public grassroots movements in the policy of space astronomy initiatives, present some recent cases which illustrate policy decisions involving broader interest groups, and consider new avenues of PE including crowdfunding and crowdsourcing.

## 1. Introduction

The role of public opinion on space policy has dramatically evolved over the last half-century. The public movements discussed here illustrate a trend away from the "technocratic" view that the public is an obstacle to progress [1]. Rather, within the framework of NASA-led space-telescope missions, a new model is forming in which the "power of the people" slowly approaches that of the government.

The relationship between science, technology, and the public is complex. The innovations resulting from space exploration, research, and development are not only generally accepted, but taken for granted, by the public [2]. Yet the public remains overwhelmingly uninformed about both the workings of NASA and the federal budget dedicated to space science. While the majority of Americans rated NASA as doing a "good" or "excellent" job during the space age, fewer than half believed the Apollo program was worth the cost [3]. Most Americans additionally overestimated the budget for NASA as a percentage of the total federal budget, and believed tax dollars could be better spent elsewhere [4]. As NASA Historian Roger Launius explains, "*the American public is notorious for its willingness to support programs in principle but to oppose their funding at levels appropriate to sustain them*" [3]. While still generally true, in recent years support by the public has grown markedly more vocal. Public interest in the Hubble and James Webb Space Telescopes surpassed casual levels of curiosity – a dedicated following of space enthusiasts and citizen scientists went so far as to directly challenge the government when both telescopes faced cancellation.

Even the opposition to funding is slowly fading as is illustrated by the public adoption of the ISEE-3 mission, due in part to the rising popularity of crowdfunding for space exploration. Crowdfunding presents the public with a unique opportunity to directly impact real research and personally benefit through rewards given at various funding levels.

The three cases discussed here epitomize this evolving dynamic between space astronomy research, the public, and the government.

## 2. Hubble Space Telescope

In the late 1960s the astronomy community in the United States recognized the need for a large telescope positioned above Earth's atmosphere and thus plans for the Large Space Telescope were born. Later renamed after the renowned astronomer Edwin Hubble, the Hubble Space Telescope experienced a series of setbacks before attaining the celebrity status it enjoys today [5]. Particularly in Hubble's later years, the role of the public became instrumental in securing the telescope's continued success.

In 1975, the House of Representatives Appropriations Subcommittee originally denied funding for Hubble due to the project's large ticket price of $500 million [6]. Between lobbying by the astronomical community and a budget reduction of almost 50% through collaboration with the European Space Agency (ESA), NASA was able to secure funding for the telescope by 1977. As a result of a series of launch delays and the Space Shuttle *Challenger* accident in 1986, the telescope waited another four years before deployment into low Earth orbit by the Space Shuttle *Discovery* in 1990 [5].

When news of a spherical aberration in Hubble's main mirror reached the media in 1990, Hubble was painted as a failure and national embarrassment [7]. After the first servicing mission to replace the defective mirror, NASA launched a highly successful nationwide outreach campaign to rehabilitate Hubble's reputation in the public eye. The Hubble Heritage Project, beginning in 1998, oversaw the online dissemination of some of Hubble's more spectacular images of the cosmos [8]. Such images have been instrumental in increasing public awareness and interest in the fields of astronomy and astrophysics.

In the year following the 2003 *Columbia* explosion, NASA Administrator Sean O'Keefe announced the decision to cancel the fifth servicing flight to Hubble, citing the mission as too risky [9,10]. Without the necessary repairs, Hubble, and all scientific research reliant on the instrument, faced a bleak future.

The public, who originally mocked the faulty instrument, emerged to fight for the telescope through a public grassroots movement of "Hubble Huggers" [11]. Numerous online petitions highlighted the ways in which abandonment of Hubble would not only damage America's future as a scientific powerhouse but also tarnish public pride and interest in astronomy [11]. NASA opened an unprecedented direct dialogue with the public to explore options to keep the telescope in orbit and operational [11,12,13]. As national concern for the fate of "the people's telescope" reached an all-time high, the American Astronomical Society and bipartisan efforts by the US Senate successfully revived the servicing mission [14, 15,16]. In 2006 Michael Griffin, who succeeded O'Keefe as NASA Administrator, announced to a standing ovation that a manned servicing mission would happen in 2009, leaving the telescope functional beyond 2014 and possibly into 2020 [5,14,17].

The public's instrumental role in saving the telescope was the focus of the 2012 documentary, *Saving Hubble* [14,18]. The film's premiere was well timed, as another social movement echoing that which saved Hubble had just taken root. The next generation of Hubble Huggers, united under the moniker "Science Warriors," launched a campaign to save another instrument - the James Webb Space Telescope.

## 3. James Webb Space Telescope

In 1996, following the successful mission to correct Hubble's mirror and the long-awaited public release of Hubble's spectacular images of space, NASA turned its sights towards the future of space telescopes. The Academy of Sciences National Research Council crowned the then-called Next Generation Space Telescope as the top scientific priority of the 2001 Astronomy Decadal Survey [19].

The telescope, renamed the James Webb Space Telescope (JWST) after the Apollo-era NASA Administrator, was originally projected for launch in 2007. A series of delays pushed back the launch date by eleven years and the estimated total cost ballooned from $0.5 million in 1997 to $8.7 million in 2011, prompting the US House of Representatives Committee on Appropriations to end all funding for the telescope for fiscal year 2012 [20,21]. Massive overspending and "poor management" were cited as reasons to completely defund the project and send a strong message to NASA that there would be "clear consequences for failing to meet... expectations" [18].

This series of budget increases and launch delays risked JWST's credibility in the public eye, just as Hubble's defective mirror led the media to question NASA's competence. Of particular concern to the astronomical community was that JWST not amass negative attention as the futures of both the telescope and subsequent space astronomy missions were perceived to be at stake. The American Astronomical Society and The Planetary Society released statements in defence of JWST, setting in motion a wave of public support as the issue gained visibility in the press [22].

Reminiscent of the "Hubble Huggers" effort in 2004, a new, largely internet-led movement of "Science Warriors" [23] voiced their dismay at the project cancellation. Social media users, employing the hashtag #saveJWST, launched movements on Facebook, Twitter, Change.org, and numerous blogs and forums to remind elected officials that the public wanted JWST to succeed [24,25,26]. A large facet of the movement focused on spurring a large-scale letter-writing campaign to government representatives [23]. The efforts proved successful when a Senate Panel voted to restore funding for the JWST, allowing the telescope to continue development with a current expected launch date of 2018 [27].

## 4. New Trends: ISEE-3 and Crowdfunding as a Public Support Tool

The influence of the public on astronomy and space exploration extends beyond serving as a mechanism to ensure financial backing from the government. Sufficient public interest in a particular space mission can entirely replace the government as the funding agent, as has recently been witnessed.

In 1978, NASA launched the International Sun-Earth Explorer 3 (ISEE-3) spacecraft, the third satellite in a NASA, ESA, and European Space Research Organisation collaboration to study solar winds and the Earth's magnetic field [28]. After being repurposed in 1985 to execute the world's first encounter with a comet and subsequently renamed the International Cometary Explorer, the spacecraft retired to make the almost 30-year journey around the sun [29].

ISEE-3 remained operational and broadcasted continually during its orbit around the Sun, prompting NASA to consider reviving contact in 2014 as it became clear the satellite could still perform scientific research [29]. Despite the opportunity, NASA deemed the cost to resuscitate the spacecraft too large to justify the project, prompting the public and astronomical community to intercede.

To save their respective telescopes, the Hubble Huggers and Science Warriors focused on convincing government representatives to restore federal funding to NASA. However, successfully saving ISEE-3 was contingent on the citizen scientists' and space enthusiasts' willingness to completely take over the space mission.

Based out of an abandoned McDonald's building, a "Reboot Team" of citizen scientists working in partnership with private space company Skycorp approached NASA with a proposal to resurrect the spacecraft [29]. With NASA's cooperation, they planned to redirect ISEE-3 to the Lagrangian 1 stable orbit point between the Earth and Sun [28]. Once there, the probe would resume its original 1978 mission goal to study the flow of solar wind from the Sun [29,30]. In May 2014, NASA issued an unprecedented Non-Reimbursable Space Act Agreement with Skycorp to hand over the space probe to the Reboot Team [29,31].

The mission was not without challenges. NASA abandoned the capability for communication with the spacecraft in its 1999 decision to upgrade the transmitters on the Deep Space Network of radio telescopes. However, in May 2014, the Reboot Team acquired the necessary hardware and worked with the Arecibo Observatory in Puerto Rico to communicate with ISEE-3. Upon successfully redirecting the probe, a large tracking antenna and radio telescope at Morehead State University would be used for the rest of the mission, with extra support provided by the Bochum Observatory in Germany [32].

NASA additionally would not provide any funding toward the mission. Consequently, the Reboot Team launched a social media campaign to increase awareness of the unique opportunity presented to citizen scientists. With the support of space enthusiasts online, they raised nearly $160,000 for the mission through crowdfunding - $35,000 more than the original goal [30].

On July 2, 2014, the Reboot Team successfully fired the ISEE-3's thrusters for the first time since 1987 [33]. Six days later, however, they were unable to adjust the trajectory of the probe to enter Earth orbit [34]. On July 10, the team turned to the public for help, stating on the *Space College* website, "we have a crowdsourced research project for our ISEE-3 Reboot fans" [35]. The call for assistance caught the attention of professional propulsion engineers at large aerospace firms who offered their expertise [36]. The team resolved to let ISEE-3 resume its original orbit around the Sun and instead focus entirely on their citizen science objectives, which remained unaffected by the probes trajectory. The ISEE-3 Interplanetary Citizen Science Mission begins on August 10 as the first "citizen science, crowd funded, crowd sourced, interplanetary space science mission" [37].

As in the previous cases, blogs and social media were instrumental in amassing support for the project, and perhaps most notably reaching a public largely unfamiliar with movements supporting space science. The Reboot Team estimated that over 95% of donors to the crowdfunding campaign were "new to overt participation in a public space effort" [38].

The successful movement to save ISEE-3 is undoubtedly an accomplishment for citizen scientists and the public. The Space Act Agreement between NASA and Skycorp, the first of its kind in which a government-funded spacecraft was handed over to citizen scientists and the public, epitomizes the evolving way in which the government and public participate in and share in space science [29]. Mirroring the Reboot Team's successful crowdfunding campaign, space enthusiasts elsewhere are adopting crowdfunding as a strategy for saving space science instruments, as is shown in Section 5.

## 5. Looking Ahead: Emerging Cases

Beyond these three examples, early 2014 has seen a growing number of public engagement initiatives to directly involve the public in policy issues.

The University of California Office of the President announced its intention to terminate funding for the ground-based Lick Observatory by 2018 [39,40]. Often it is the case that a previous movement inspires the next, revealing the interconnectivity of these communities. Just as the Hubble Huggers set a precedent for the saveJWST Science Warriors, the recently emergent Save Lick Observatory movement benefited from support by the Science Warriors [24]. The Save Lick campaign to keep the observatory alive is simultaneously challenging the university's decision and raising its own money through private donations [41].

A growing number of international examples following this trend have also surfaced, including Canada's first space telescope mission, the Microvariability and Oscillations of STars (MOST) Telescope. MOST will be retired due to funding cuts to the Canadian Space Agency [42]. Like ISEE-3, MOST is still fully operational, sparking discussions of crowdfunding as a potential strategy to sustain the telescope [43].

## 6. Conclusions

These examples illustrate that public opinion and organized social movements do influence space policy and government funding. The trends explored here suggest that in recent years there have indeed been changes in the government-public-science dynamic, which lend themselves, upon further investigation, to a new model of understanding these relations. Social media and blogs have revolutionized space astronomy communications both between the government and the public, and among the various communities of space enthusiasts. The advent of crowdfunding has additionally opened new venues through which the public can directly engage with, and change, space policy. At present we must rely solely on anecdotal evidence from specific cases; additional work and data are necessary to fully understand the mechanisms of public opinion and grassroots movement on space-related policies.

## Acknowledgements
This research was done within the framework of the Leiden/ESA Astrophysics Program for Summer Students hosted by Leiden Observatory and European Space Agency. We would like to thank George Miley, Jos van den Broek, Sophie A. Jones, Dilovan Serindag, and S. R. J. Walton for their comments on this article.